\documentclass[aps,prb,reprint,superscriptaddress,showkeys,preprintnumbers]{revtex4-2}
\usepackage[T1]{fontenc}
\usepackage{float}
\usepackage{times}
\usepackage{graphicx,color}
\usepackage{amsfonts,amsmath,amssymb,amsbsy}
\usepackage[colorlinks=true, linkcolor=blue, citecolor=blue, urlcolor=blue]{hyperref}
\renewcommand{\figurename}{\textbf{FIGURE}}

\def\section#1{\medskip\noindent\textbf{\large{#1}}\par}
\def\subsection#1{\medskip\noindent\textbf{\large{#1}}\par}
\makeatletter
\renewcommand{\fnum@figure}{\figurename~\textbf{\thefigure}}
\makeatother
\def\blue#1{\textcolor{blue}{#1}}

\def\emph#1{\textcolor{blue}{#1}}
\def\emph#1{\textcolor{black}{#1}}
\begin{document}
%%%%%%%%%%%%%%%%%%%%%%%%%%%%%%%%%%%%%%%%%%%%%%%%%%%%%%%%%%%%

\title{Bimerons create bimerons: proliferation and aggregation induced by \emph{currents and magnetic fields}}

\author{Xichao Zhang}
%\email[Email:~]{xichaozhang@aoni.waseda.jp}
\affiliation{Department of Applied Physics, Waseda University, Okubo, Shinjuku-ku, Tokyo 169-8555, Japan}

\author{Yan Zhou}
%\email[Email:~]{zhouyan@cuhk.edu.cn}
\affiliation{School of Science and Engineering, The Chinese University of Hong Kong, Shenzhen, Guangdong 518172, China}

\author{Xiuzhen Yu}
%\email[Email:~]{yu_x@riken.jp}
\affiliation{RIKEN Center for Emergent Matter Science (CEMS), Wako 351-0198, Japan}

\author{Masahito Mochizuki}
\email[Email:~]{masa_mochizuki@waseda.jp}
\affiliation{Department of Applied Physics, Waseda University, Okubo, Shinjuku-ku, Tokyo 169-8555, Japan}

%-%-%-%-%-%-%-%-%-%-%-%-%-%-%-%-%-%-%-%-%-%-%-%-%-%-%-%-%-%-%
\begin{abstract}
The aggregation of topological spin textures at nano and micro scales has practical applications in spintronic technologies. Here, the authors report the in‐plane current‐induced proliferation and aggregation of bimerons in a bulk chiral magnet. It is found that the spin‐transfer torques can induce the proliferation and aggregation of bimerons only in the presence of an appropriate out‐of‐plane magnetic field. It is also found that a relatively small damping and a relatively large non‐adiabatic spin‐transfer torque could lead to more pronounced bimeron proliferation and aggregation. Particularly, the current density should be larger than a certain threshold in order to trigger the proliferation; namely, the bimerons may only be driven into translational motion under weak current injection. Besides, the authors find that the aggregate bimerons could relax into a deformed honeycomb bimeron lattice with a few lattice structure defects after the current injection. The results are promising for the development of bio‐inspired spintronic devices that use a large number of aggregate bimerons. The findings also provide a platform for studying aggregation‐induced effects in spintronic systems, such as the aggregation‐induced lattice phase transitions.
\end{abstract}
%-%-%-%-%-%-%-%-%-%-%-%-%-%-%-%-%-%-%-%-%-%-%-%-%-%-%-%-%-%-%

%\date{\today}
\date{July 10, 2024}

\preprint{\hyperlink{https://doi.org/10.1002/agt2.590}{DOI: 10.1002/agt2.590}}


\maketitle

%-%-%-%-%-%-%-%-%-%-%-%-%-%-%-%-%-%-%-%-%-%-%-%-%-%-%-%-%-%-%
%Memo: Aggregation means the formation of a number of things into a cluster.
%-%-%-%-%-%-%-%-%-%-%-%-%-%-%-%-%-%-%-%-%-%-%-%-%-%-%-%-%-%-%

%-%-%-%-%-%-%-%-%-%-%-%-%-%-%-%-%-%-%-%-%-%-%-%-%-%-%-%-%-%-%
\section{1 | INTRODUCTION}
\label{se:Introduction}
%-%-%-%-%-%-%-%-%-%-%-%-%-%-%-%-%-%-%-%-%-%-%-%-%-%-%-%-%-%-%

%%%%%%%%%%%%%%%%%%%%%%%%%%%%%%%%%%%%%%%%%%%%%%%%%%%%%%%%%%%%
\begin{figure*}[t]
\centerline{\includegraphics[width=0.85\textwidth]{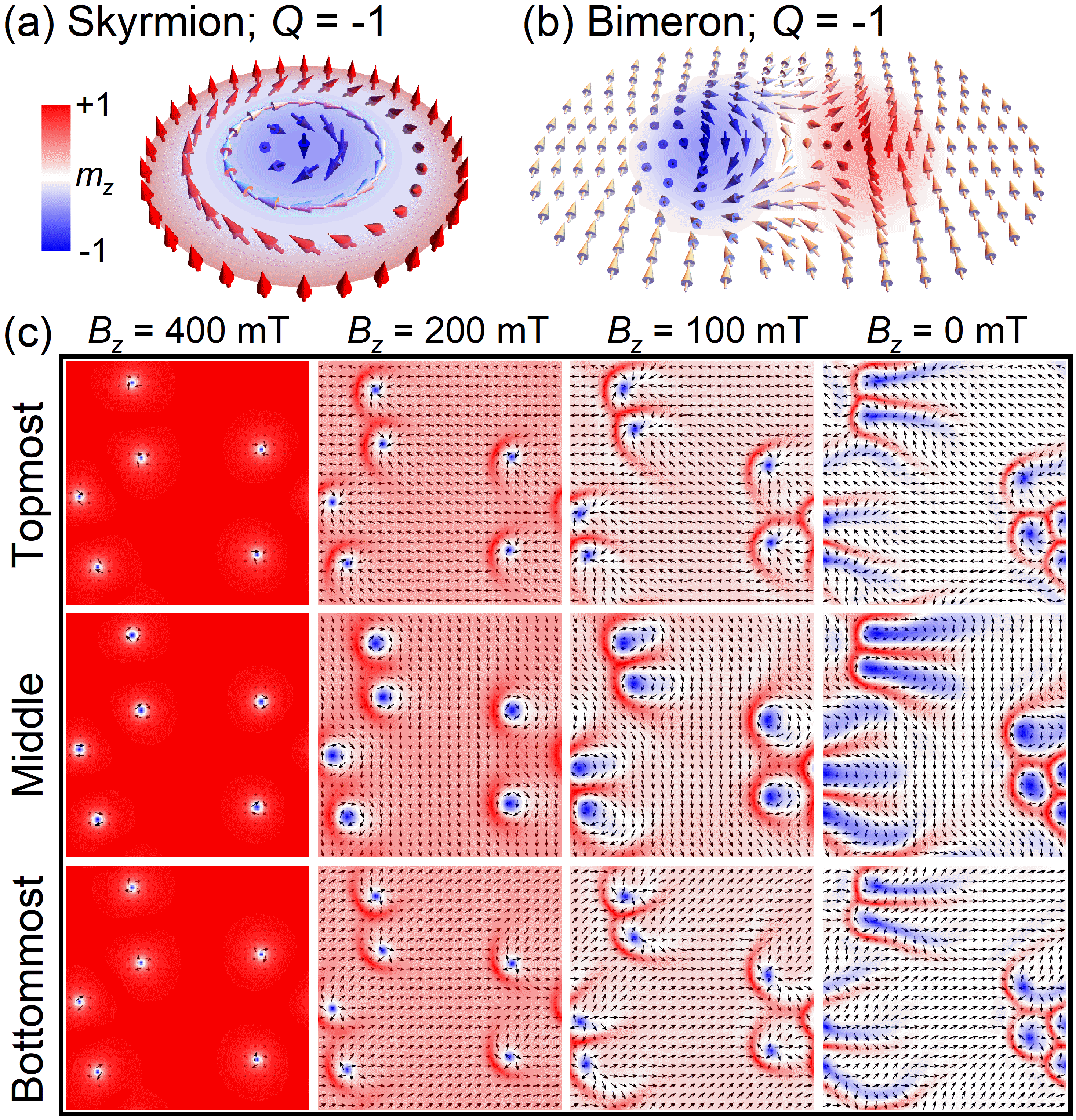}}
\caption{%
The initial spin configurations at different external magnetic fields.
(A) Schematic illustration of a standard Bloch-type skyrmion with $Q=-1$ that can be found in the bulk magnet with chiral exchange interactions. The color scale represents the $m_z$ component, which has been used throughout the paper.
(B) Schematic illustration of a standard Bloch-type bimeron with $Q=-1$ in the bulk magnet with chiral exchange interactions. The bimeron is a topological counterpart to the skyrmion as they carry the same topological charge; however, the bimeron has two out-of-plane cores while the skyrmion has only one.
(C) The spin configurations in the bulk magnet at different external magnetic fields. An out-of-plane magnetic field $B_z$ is applied to the sample with a randomly distributed initial spin configuration. The magnetic field first increases from $B_z=0$ mT to $400$ mT with a step change of $10$ mT, and then decreased to $0$ mT with the same step change. The system is relaxed before each change of the magnetic field. The top views of the spin configurations at $B_z=400$ mT, $200$ mT, $100$ mT, and $0$ mT during the decrease of the magnetic field are given. These field-dependent states, including bimerons or skyrmions, will be used in the current-driven dynamics simulation as the initial states. The thickness of the bulk magnet equals $80$ nm. The top views focus on the spins in the topmost, middle, and bottommost horizontal layers of the bulk magnet.
}
\label{FIG1}
\end{figure*}
%%%%%%%%%%%%%%%%%%%%%%%%%%%%%%%%%%%%%%%%%%%%%%%%%%%%%%%%%%%%

%%%%%%%%%%%%%%%%%%%%%%%%%%%%%%%%%%%%%%%%%%%%%%%%%%%%%%%%%%%%
\begin{figure*}[t]
\centerline{\includegraphics[width=0.99\textwidth]{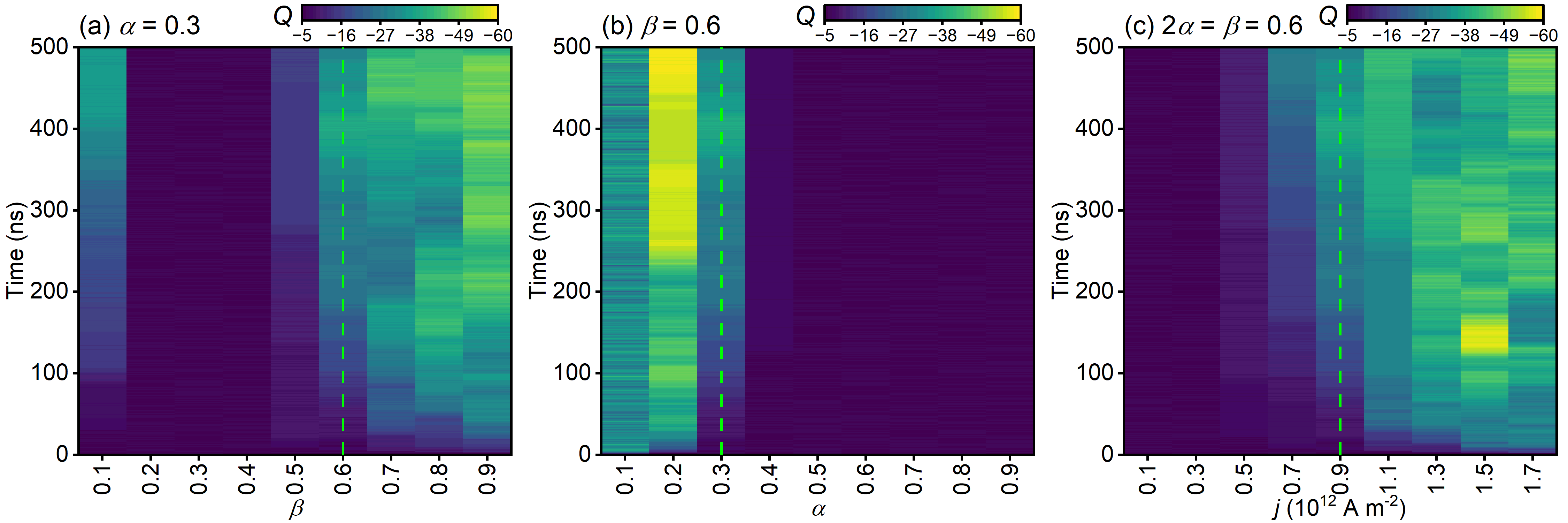}}
\caption{%
Current-driven proliferation and aggregation of bimerons in the presence of an out-of-plane magnetic field of $B_z=+100$ mT.
(A) Total topological charge $Q$ as functions of time $t$ and non-adiabatic spin-transfer torque strength $\beta$. Here, an in-plane current with the current density of $j=0.9\times 10^{12}$ A m$^{-2}$ is applied to drive the bimerons in the model with a damping parameter of $\alpha=0.3$.
(B) $Q$ as functions of $t$ and $\alpha$. Here, an in-plane current with $j=0.9\times 10^{12}$ A m$^{-2}$ is applied to drive the bimerons in the model with $\beta=0.6$.
(C) $Q$ as functions of $t$ and $j$. Here, an in-plane current is applied to drive the bimerons in the model with $\alpha=0.3$ and $\beta=0.6$.
Note that the total topological charge of the initial state obtained at $B_z=+100$ mT equals $-6$ at $t=0$ ns [see Figure~\ref{FIG1}(C)]. The green dashed lines indicate the model with the same parameters (i.e., $\alpha=0.3$, $\beta=0.6$, and $j=0.9\times 10^{12}$ A m$^{-2}$), which is the focused case in this work.
}
\label{FIG2}
\end{figure*}
%%%%%%%%%%%%%%%%%%%%%%%%%%%%%%%%%%%%%%%%%%%%%%%%%%%%%%%%%%%%

\noindent
Topological structures can be found everywhere in different forms and usually have important implications on different matter~\cite{Wieder_2022,Bernevig_2022,Yang_2021,Gilbert_2021,Braun_2012,Ozawa_2019,Shen_2024,Shankar_2022,Tobak_1982,Moffatt_2021}.
For example, the blood flow with topologically non-trivial vortex structures inside the human heart is found to play a crucial role in an efficient blood supply from the heart to organs~\cite{Sakajo_2023}.
The vortex structure is the most famous spiral structure in nature that carries a half integer topological charge~\cite{DeAlfaro_PLB1976,Antos_2008,Kosterlitz_2017,Behncke_2018,Shen_2019,Zhang_JPCM2020}.
A recent study also suggests that the vortex-like rotational wave patterns are widespread during both resting and cognitive task states in the human brain~\cite{Xu_2023}.

In magnetic materials, it is also possible to find unique topological structures, such as the vortex and meron structures~\cite{Ezawa_PRB2011,Lin_PRB2015,Kharkov_PRL2017,Leonov_PRB2017,Yu_Nature2018,Chmiel_NM2018,Kolesnikov_SR2018,Fernandes_SSC2019,Gobel_PRB2019,Moon_PRApplied2019,Murooka_SR2020,Shen_PRL2020,Lu_2020,ES_PRB2020,Wu_NE2023,Yu_AM2024,Bhukta_NC2024}.
Both the vortex and meron structures carry a half integer topological charge, and their order parameters away from their cores lie in the $x$-$y$ plane of the magnetic films. The order parameters inside their cores smoothly rotate to the out-of-plane directions. The difference between a vortex and a meron is that the in-plane spins of a meron form a pure radial configuration, while the in-plane spins of a vortex is usually of a clockwise or counterclockwise curling configuration~\cite{Moon_PRB1995}.
However, in magnets with chiral exchange interactions, i.e., the Dzyaloshinskii-Moriya (DM) interactions~\cite{Dzyaloshinsky,Moriya}, a pair of merons or vortices carrying an integer topological charge of $Q=\pm 1$ is usually more stable.
The paired merons or vortices with $Q=\pm 1$ in in-plane magnetized systems are also called bimerons~\cite{Ezawa_PRB2011,Lin_PRB2015,Kharkov_PRL2017,Leonov_PRB2017,Yu_Nature2018,Chmiel_NM2018,Kolesnikov_SR2018,Fernandes_SSC2019,Gobel_PRB2019,Moon_PRApplied2019,Murooka_SR2020,Shen_PRL2020,Lu_2020,ES_PRB2020,Wu_NE2023,Yu_AM2024,Moon_PRB1995,Ezawa_2010,Ezawa_2011,Silva_PRB2014,Kim_PRB2019,Zhang_PRB2020,Li_NJPCM2020,Silva_PRB2020,Shen_PRB2020,Sun_NC2020,Jani_N2021,Nagase_NC2021,Zhang_APL2021,Zhang_PRB2021,Shen_PRB2022,Ohara_NL2022,Jin_NJP2022,Wang_JMMM2022,Silva_PLA2022,Castro_PRB2023,Liang_PRB2023,Hu_JPDAP2023,Babu_2024,Bhukta_NC2024}, which are topological counterparts of skyrmions with $Q=\pm 1$ in perpendicularly magnetized systems~\cite{Nagaosa_NNANO2013,Mochizuki_JPCM2015,Wiesendanger_NATREVMAT2016,Finocchio_JPD2016,Fert_NATREVMAT2017,Wanjun_PHYSREP2017,Kanazawa_AM2017,Everschor_JAP2018,Back_JPD2020,Fujishiro_APL2020,Gobel_PHYSREP2021,Reichhardt_RMP2022,Bogdanov_SPJETP1989,Roszler_NATURE2006,Muhlbauer_SCIENCE2009,Yu_NATURE2010,Bogdanov_NRP2020,Zhang_JPCM2020}.

Both magnetic bimerons and skyrmions are promising particle-like bit-carrying candidates for future in-memory and bio-inspired computing technologies because of their non-volatility, radiation hardness, and low energy consumption~\cite{Kang_PIEEE2016,Li_MH2021,Luo_APLM2021,Marrows_APL2021,Vakili_JAP2021,Li_SB2022,Zhang_JPCM2020}.
They can respond to different external stimuli and show interesting dynamic phenomena that are fundamental to practical applications~\cite{Ezawa_2010,Ezawa_2011,Silva_PRB2014,Kim_PRB2019,Zhang_PRB2020,ES_PRB2020,Li_NJPCM2020,Silva_PRB2020,Shen_PRB2020,Sun_NC2020,Jani_N2021,Nagase_NC2021,Zhang_APL2021,Zhang_PRB2021,Shen_PRB2022,Ohara_NL2022,Jin_NJP2022,Wang_JMMM2022,Silva_PLA2022,Castro_PRB2023,Liang_PRB2023,Hu_JPDAP2023,Babu_2024,Kang_PIEEE2016,Li_MH2021,Luo_APLM2021,Marrows_APL2021,Vakili_JAP2021,Li_SB2022,Zang_PRL2011,Wanjun_NPHYS2017,Litzius_NPHYS2017,Yu_NP2018,Sampaio_NN2013,Tomasello_SREP2014,Xichao_NC2016,Xichao_PRB2016B,Xichao_PRB2022A,Xichao_PRB2022B,Zhang_NL2023,Schott_NL2017,Ma_NL2019,Lin_PRB2013,Reichhardt_PRB2015A,Reichhardt_NJP2016,Reichhardt_JPCM2019,Reichhardt_PRB2019,Reichhardt_PRB2020,Souza_PRB2021,Vizarim_PRB2022,Souza_NJP2022,Zhang_PRB2023,Rocha_JPCM2024,Souza_PRB2024}.
Their manipulatable internal structures may also be harnessed for quantum computation~\cite{Psaroudaki_PRL2021,Xia_PRL2022,Xia_CM2022,Psaroudaki_APL2023}.
In contrast to the skyrmion, which only has a single out-of-plane core [Figure~\ref{FIG1}(A)], the bimeron has more degrees of freedom as a single isolated bimeron has two out-of-plane cores and the spacing between the two cores is adjustable at certain conditions [Figure~\ref{FIG1}(B)]. Therefore, it is envisioned that the structure-dependent dynamics of bimerons could result in more complicated but intriguing phenomena.

Basic dynamic effects and phenomena of topological structures in magnets include the creation, motion, and transformation.
The motion of a single isolated bimeron driven by either spin currents~\cite{Lin_PRB2015,Gobel_PRB2019,Moon_PRApplied2019,Shen_PRL2020,ES_PRB2020,Kim_PRB2019,Zhang_PRB2020,Li_NJPCM2020,Silva_PRB2020,Shen_PRB2020,Silva_PLA2022,Hu_JPDAP2023,Babu_2024} or spin waves~\cite{Liang_PRB2023} has been explored, and some recent studies have demonstrated the transformation between bimerons and skyrmions~\cite{Zhang_PRB2021,Ohara_NL2022,Yu_AM2024,Castro_PRB2023}.
Besides, both simulations and experiments have suggested that bimerons can form clusters during the current-driven motion of isolated bimerons~\cite{Yu_AM2024,Zhang_PRB2020,Li_NJPCM2020}.
However, the underlying mechanism for the proliferation and aggregation of bimerons during the current-driven motion of bimerons has remained elusive.
The proliferation of bimerons is particularly vital for applications that require a large number of topological structures to exist in a chamber, such as the artificial synapses and neurons~\cite{Li_MH2021,Song_NE2020,Li_N2017,Huang_N2017,Lee_PRA2022,Lee_SR2023}.
The realization of aggregate bimerons may also stimulate more multidisciplinary research in the field of aggregate science~\cite{Tang_2011}.

In this work, we report the proliferation and aggregation of bimerons during the motion of bimerons driven by an in-plane current in the presence of an out-of-plane magnetic field, where many new bimerons are rapidly created and aggregate into clusters in a dynamic way.
The current-induced proliferation of bimerons is found to be a unique dynamic feature to bimerons, which means seed bimerons can create new bimerons.
Our findings also suggest that a suitable driving force setup can create numerous bimerons from a few seed bimerons, which serves as an effective way to produce topological structures in magnets.

This paper is structured as follows.
In Section~\blue{2}, we present computational modeling methods.
In Section~\blue{3.1}, we show the relaxed \emph{metastable} spin configurations at different out-of-plane magnetic fields, which are used as the initial states in the simulations of current-induced dynamics.
In Section~\blue{3.2}, we focus on the proliferation and aggregation of bimerons induced by an in-plane current in the presence of an out-of-plane magnetic field.
In Section~\blue{3.3}, we demonstrate the relaxation of aggregate bimerons after the in-plane current injection.
In Section~\blue{3.4}, we investigate the effect of the applied out-of-plane magnetic field on the current-induced dynamics.
In Section~\blue{4}, we summarize our key findings and briefly discuss potential implications for future applications.

%%%%%%%%%%%%%%%%%%%%%%%%%%%%%%%%%%%%%%%%%%%%%%%%%%%%%%%%%%%%
\begin{figure*}[t]
\centerline{\includegraphics[width=0.81\textwidth]{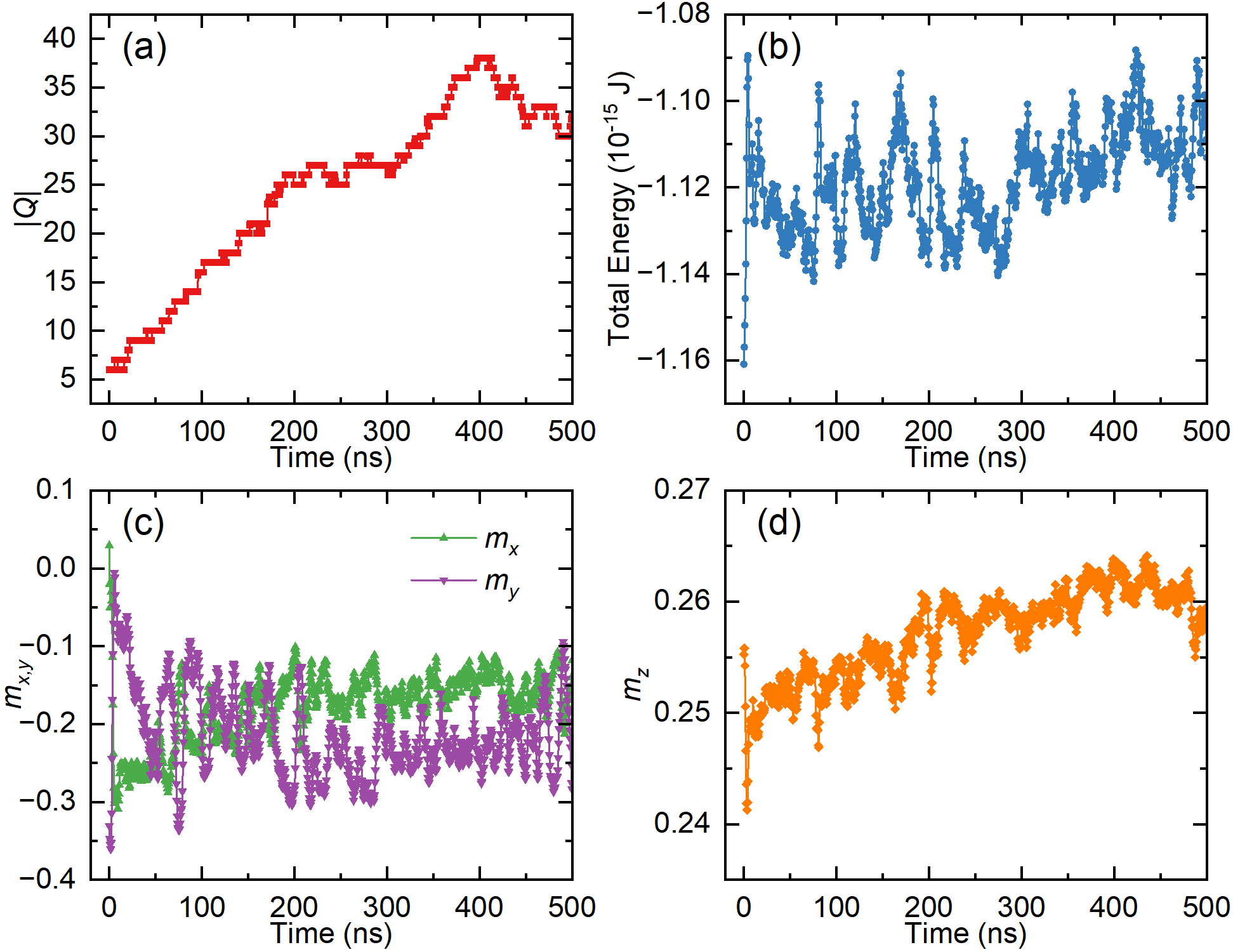}}
\caption{%
Current-driven proliferation and aggregation of bimerons in the model with $\alpha=0.3$ and $\beta=0.6$ in the presence of an out-of-plane magnetic field of $B_z=+100$ mT. An in-plane current of $j=0.9\times 10^{12}$ A m$^{-2}$ is applied to drive the dynamics during $t=0-500$ ns.
(A) Absolute total topological charge $|Q|$ of the system as a function of time.
(B) Total energy of the system as a function of time.
(C) Reduced in-plane magnetization components $m_x$ and $m_y$ as functions of time.
(D) Reduced out-of-plane magnetization component $m_z$ as a function of time.
}
\label{FIG3}
\end{figure*}
%%%%%%%%%%%%%%%%%%%%%%%%%%%%%%%%%%%%%%%%%%%%%%%%%%%%%%%%%%%%

%%%%%%%%%%%%%%%%%%%%%%%%%%%%%%%%%%%%%%%%%%%%%%%%%%%%%%%%%%%%
\begin{figure*}[t]
\centerline{\includegraphics[width=0.81\textwidth]{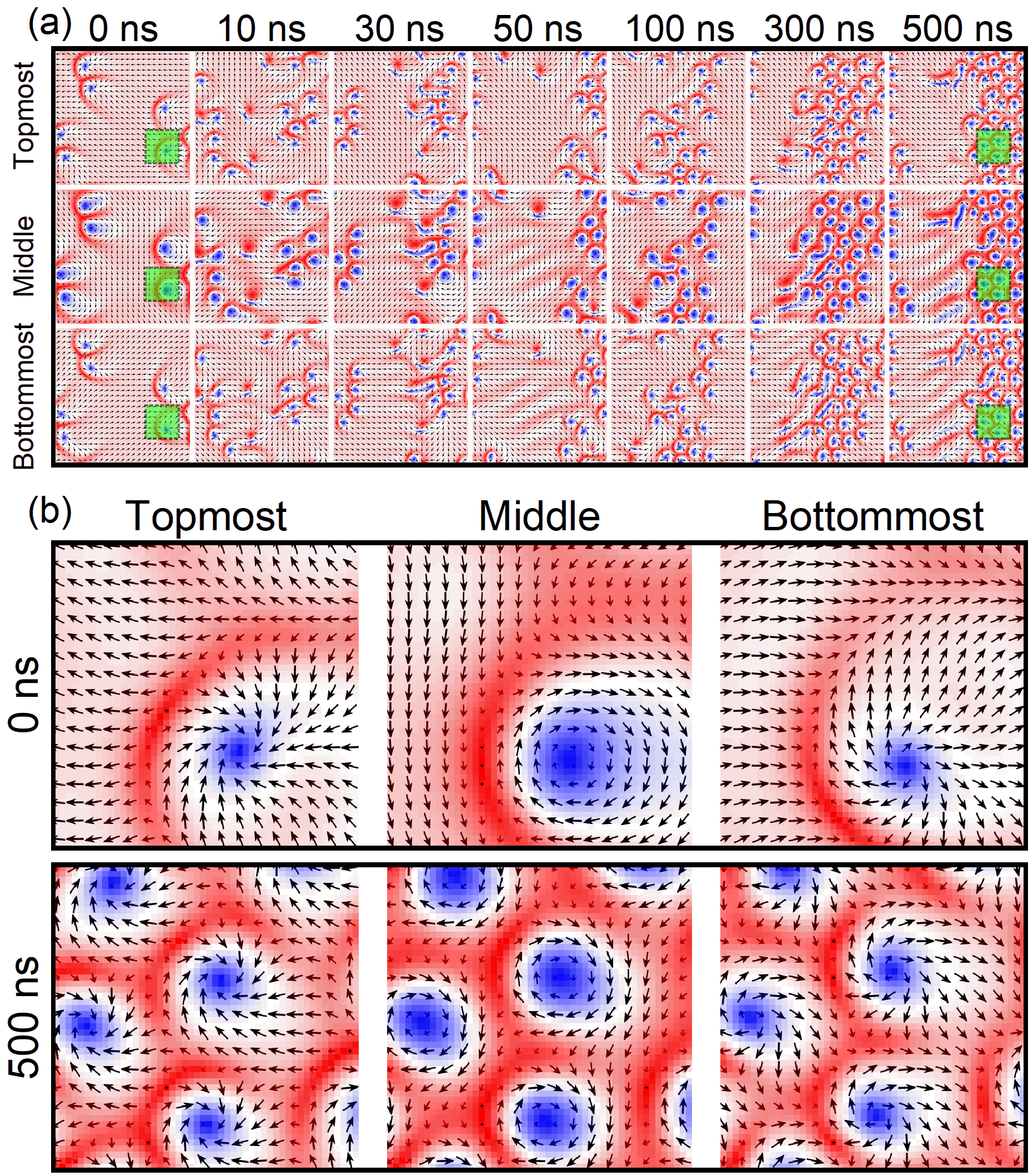}}
\caption{%
Current-driven proliferation and aggregation of bimerons in the model with $\alpha=0.3$ and $\beta=0.6$ in the presence of an out-of-plane magnetic field of $B_z=+100$ mT. An in-plane current of $j=0.9\times 10^{12}$ A m$^{-2}$ is applied to drive the dynamics during $t=0-500$ ns.
(A) Top-view snapshots of the spin configurations at selected times.
(B) Zoomed-up views of the spin configurations corresponding to the regions indicated by green boxes in (A).
}
\label{FIG4}
\end{figure*}
%%%%%%%%%%%%%%%%%%%%%%%%%%%%%%%%%%%%%%%%%%%%%%%%%%%%%%%%%%%%

%%%%%%%%%%%%%%%%%%%%%%%%%%%%%%%%%%%%%%%%%%%%%%%%%%%%%%%%%%%%
\begin{figure*}[t]
\centerline{\includegraphics[width=0.81\textwidth]{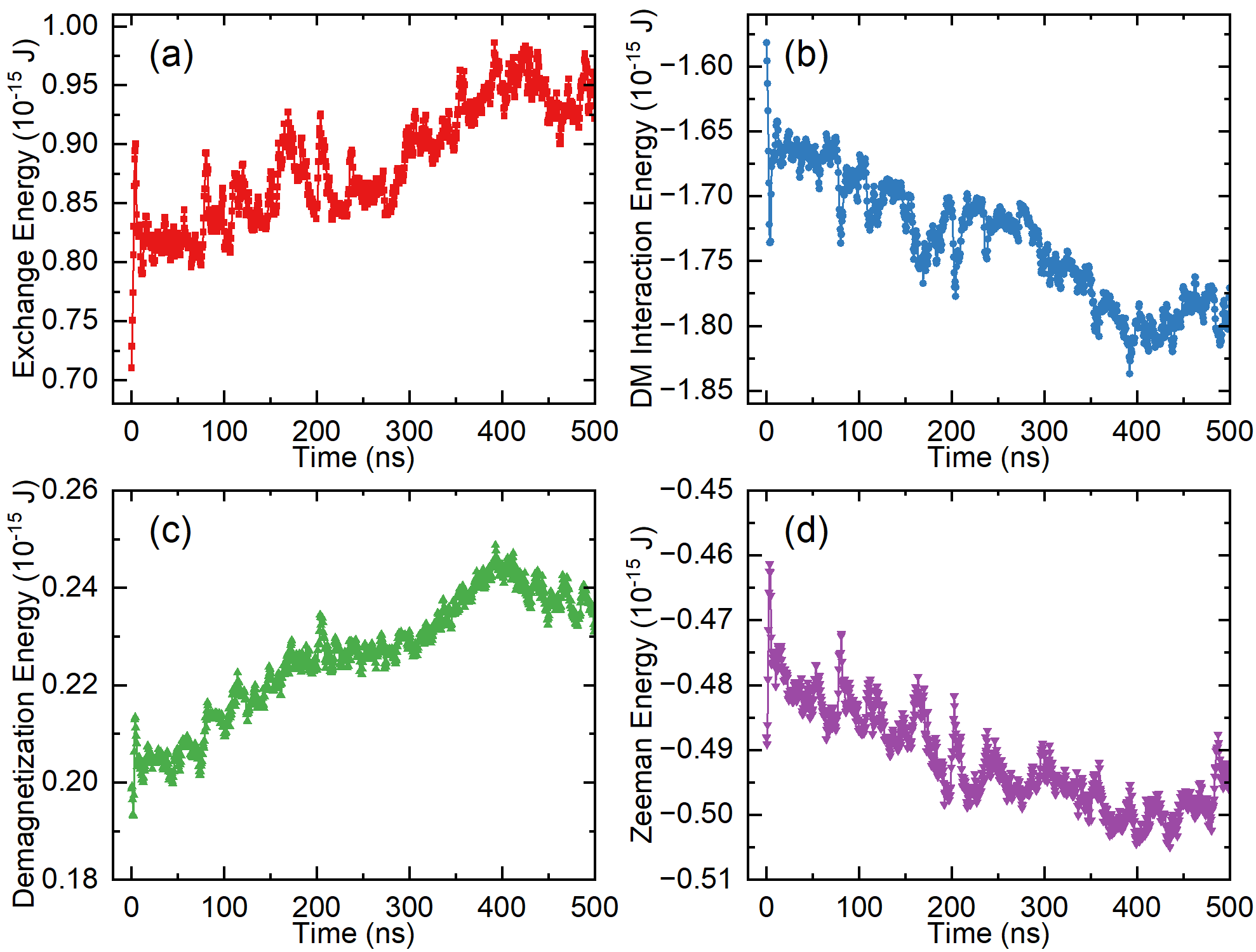}}
\caption{%
Current-driven proliferation and aggregation of bimerons in the model with $\alpha=0.3$ and $\beta=0.6$ in the presence of an out-of-plane magnetic field of $B_z=+100$ mT. An in-plane current of $j=0.9\times 10^{12}$ A m$^{-2}$ is applied to drive the dynamics during $t=0-500$ ns.
(A) Exchange energy as a function of time.
(B) Dzyaloshinskii-Moriya (DM) interaction energy as a function of time.
(C) Demagnetization energy as a function of time.
(D) Zeeman energy as a function of time.
}
\label{FIG5}
\end{figure*}
%%%%%%%%%%%%%%%%%%%%%%%%%%%%%%%%%%%%%%%%%%%%%%%%%%%%%%%%%%%%

%%%%%%%%%%%%%%%%%%%%%%%%%%%%%%%%%%%%%%%%%%%%%%%%%%%%%%%%%%%%
\begin{figure*}[t]
\centerline{\includegraphics[width=0.81\textwidth]{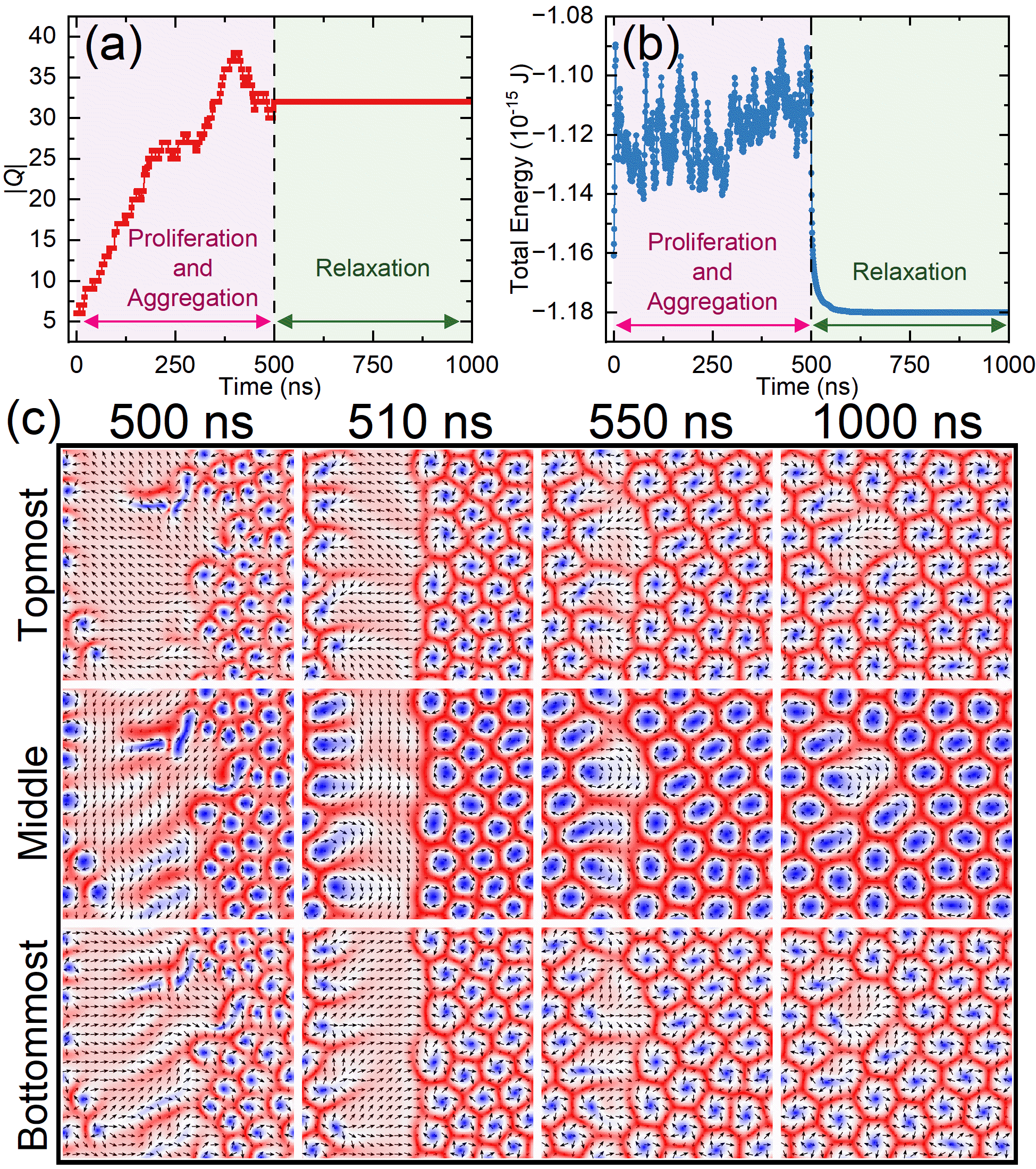}}
\caption{%
Relaxation of aggregate bimerons into a defected bimeron lattice after current-driven proliferation and aggregation of bimerons in the model with $\alpha=0.3$ and $\beta=0.6$ in the presence of an out-of-plane magnetic field of $B_z=+100$ mT.
An in-plane current of $j=0.9\times 10^{12}$ A m$^{-2}$ is applied to drive the dynamics during $t=0-500$ ns, and the system is relaxed at $j=0$ A m$^{-2}$ during $t=500-1000$ ns.
(A) Absolute total topological charge $|Q|$ of the system as a function of time.
(B) Total energy of the system as a function of time.
(C) Top-view snapshots of the spin configurations at selected times during the relaxation (i.e., $t=500-1000$ ns), showing the transition from dynamically aggregate bimerons into a static bimeron lattice with defects.
}
\label{FIG6}
\end{figure*}
%%%%%%%%%%%%%%%%%%%%%%%%%%%%%%%%%%%%%%%%%%%%%%%%%%%%%%%%%%%%

%%%%%%%%%%%%%%%%%%%%%%%%%%%%%%%%%%%%%%%%%%%%%%%%%%%%%%%%%%%%
\begin{figure}[t]
\centerline{\includegraphics[width=0.49\textwidth]{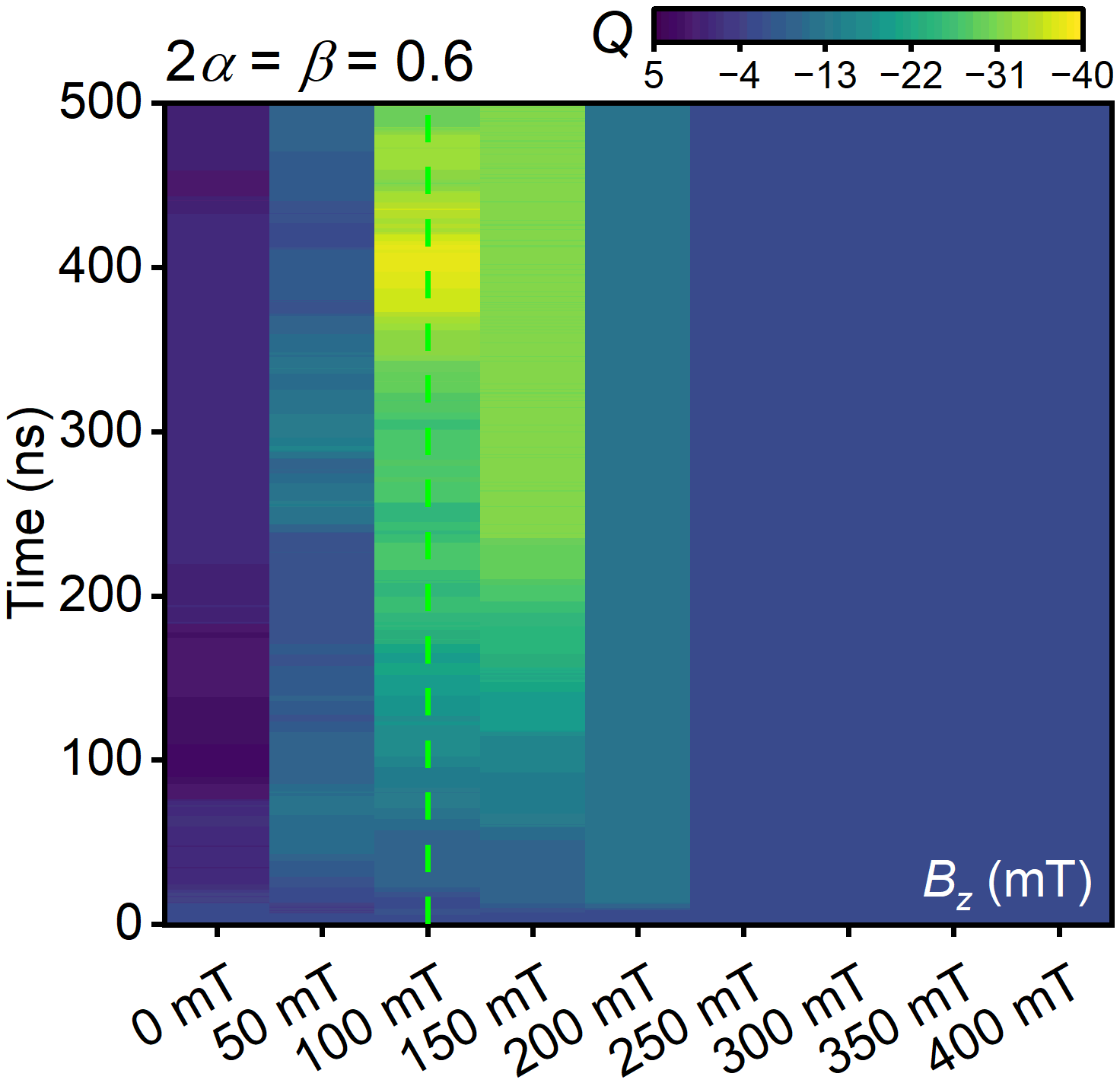}}
\caption{%
The effect of applied out-of-plane magnetic field on the current-driven proliferation and aggregation of bimerons.
Total topological charge $Q$ as functions of time $t$ and out-of-plane magnetic field $B_z$.
Here, an in-plane current with the current density of $j=0.9\times 10^{12}$ A m$^{-2}$ is applied to drive the bimerons in the model with $\alpha=0.3$ and $\beta=0.6$. The green dashed line indicates the model with the default parameters, i.e., $\alpha=0.3$, $\beta=0.6$, and $j=0.9\times 10^{12}$ A m$^{-2}$, which is the focused case in this work.
The initial states at $t=0$ ns are obtained by relaxing the model in the presence of the out-of-plane magnetic field without applying the driving current.
}
\label{FIG7}
\end{figure}
%%%%%%%%%%%%%%%%%%%%%%%%%%%%%%%%%%%%%%%%%%%%%%%%%%%%%%%%%%%%

%%%%%%%%%%%%%%%%%%%%%%%%%%%%%%%%%%%%%%%%%%%%%%%%%%%%%%%%%%%%
\begin{figure*}[t]
\centerline{\includegraphics[width=0.99\textwidth]{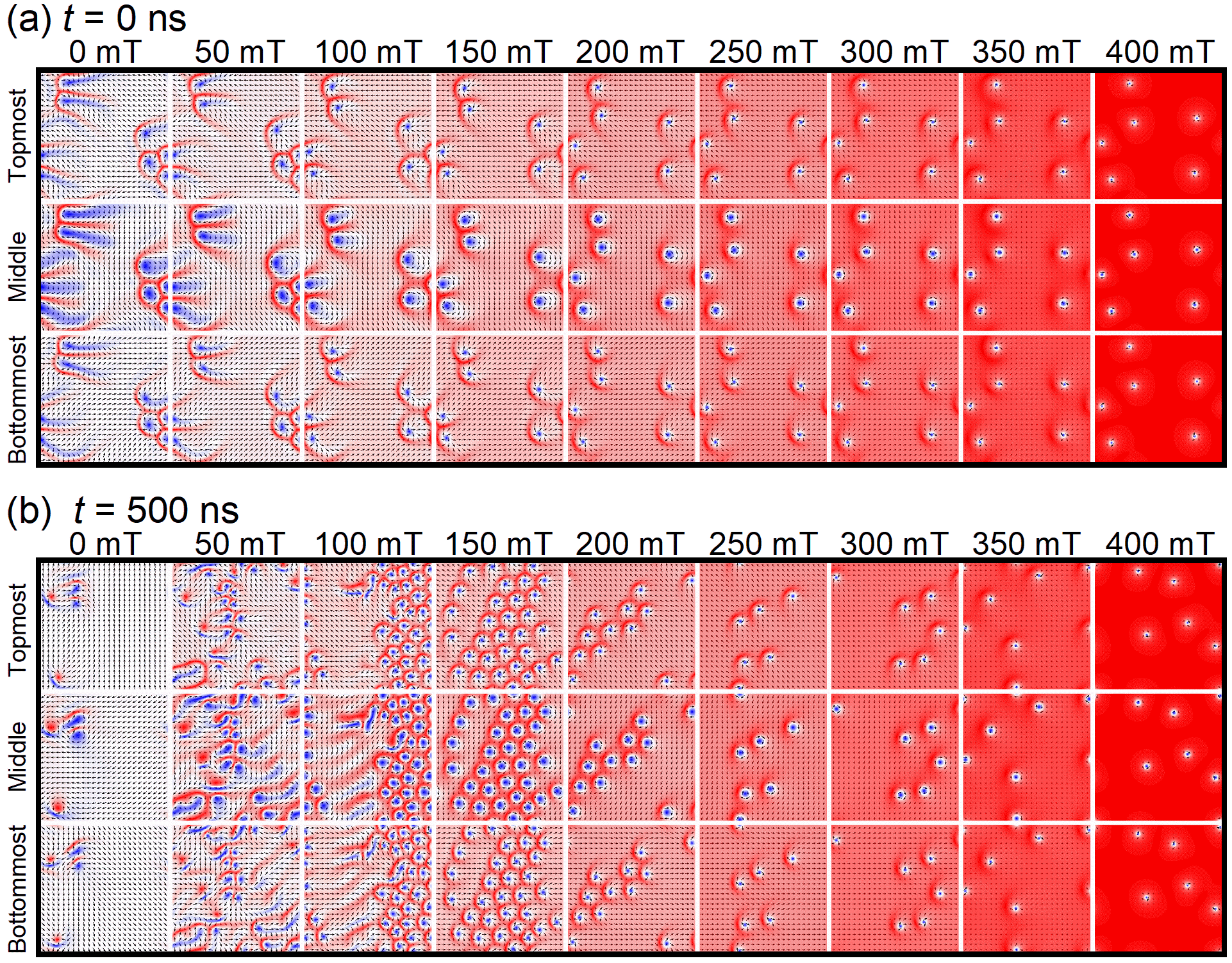}}
\caption{%
The effect of applied out-of-plane magnetic field on the current-driven proliferation and aggregation of bimerons.
Here, an in-plane current with the current density of $j=0.9\times 10^{12}$ A m$^{-2}$ is applied to drive the bimerons in the model with $\alpha=0.3$ and $\beta=0.6$.
(A) Top-view snapshots of the spin configurations of the initial states ($t=0$ ns) at different applied out-of-plane magnetic fields, i.e., $B_z=0-400$ mT.
(B) Top-view snapshots of the spin configurations of the final states ($t=500$ ns) at different applied out-of-plane magnetic fields, i.e., $B_z=0-400$ mT.
}
\label{FIG8}
\end{figure*}
%%%%%%%%%%%%%%%%%%%%%%%%%%%%%%%%%%%%%%%%%%%%%%%%%%%%%%%%%%%%

\vbox{}
%-%-%-%-%-%-%-%-%-%-%-%-%-%-%-%-%-%-%-%-%-%-%-%-%-%-%-%-%-%-%
\section{2 | SIMULATION METHODS}
\label{se:Methods}
%-%-%-%-%-%-%-%-%-%-%-%-%-%-%-%-%-%-%-%-%-%-%-%-%-%-%-%-%-%-%

\noindent
In this work, we focus on the current-driven dynamics of bimerons in a bulk magnet with chiral exchange interactions, where the existence of bimerons has been confirmed in a recent experiment~\cite{Yu_AM2024}.
The chiral exchange interactions in the bulk magnet could stabilize Bloch-type topological spin textures, including both bimerons and skyrmions~\cite{Yu_AM2024}.
The length, width, and thickness of the bulk magnet considered in our simulations are set to $1000$ nm, $1000$ nm, and $80$ nm, respectively.
We also apply the periodic boundary conditions in the $x$ and $y$ directions, so that we will be able to study the current-driven dynamics of bimerons without interacting with the boundaries.
\emph{We note that the boundary condition is an important factor to consider. We apply the periodic boundary conditions in the $x$ and $y$ dimensions to mimic the situation of a very large sample in real experiments. The sample size could be larger than tens of microns in experiments so that it is possible to study the bimeron dynamics in the center area of the real sample. If we replace the periodic boundary conditions with the open boundary conditions in the simulation, most likely the bimerons will be driven to the sample edge soon after the application of the current, which will lead to the annihilation of bimerons.}

The magnetization dynamics in the bulk magnet is controlled by the Landau-Lifshitz-Gilbert (LLG) equation augmented with the adiabatic and non-adiabatic spin-transfer torques~\cite{MuMax1,MuMax2,Zhang_IEEE2017}.
The spin-transfer torques are generated when an in-plane electric current is injected into the bulk magnet, which has been demonstrated in experiments~\cite{Yu_AM2024}.
The magnetization dynamics equation is given as
\begin{equation}
\label{eq:LLGSTT}
\begin{split}
\partial_{t}\boldsymbol{m}=&-\gamma_{0}\boldsymbol{m}\times\boldsymbol{h}_{\text{eff}}+\alpha(\boldsymbol{m}\times\partial_{t}\boldsymbol{m}) \\
&+u(\boldsymbol{m}\times\partial_{x}\boldsymbol{m}\times\boldsymbol{m})-\beta u(\boldsymbol{m}\times\partial_{x}\boldsymbol{m}),
\end{split}
\end{equation}
where $\boldsymbol{m}$ is the reduced magnetization,
$t$ is the time,
$\gamma_0$ is the absolute gyromagnetic ratio, and
$\alpha$ is the Gilbert damping parameter.
The adiabatic spin-transfer torque coefficient is given by
$u=\left|\left(\gamma_{0}\hbar/\mu_{0}e\right)\right|\cdot\left(jP/2M_{\text{S}}\right)$,
and $\beta$ is the strength of the non-adiabatic spin-transfer torque.
$\mu_{0}$ and $M_{\text{S}}$ denote the vacuum permeability constant and the saturation magnetization, respectively.
$\hbar$ is the reduced Planck constant, $e$ is the electron charge, $j$ is the applied current density, and $P$ is the spin polarization rate.
The effective field is given as
$\boldsymbol{h}_{\rm{eff}}=-\frac{1}{\mu_{0}M_{\text{S}}}\cdot\frac{\delta\varepsilon}{\delta\boldsymbol{m}}$,
where $\varepsilon$ denotes the energy density.
The energy density terms include the ferromagnetic exchange interaction, chiral exchange interaction, demagnetization, and applied magnetic field terms~\cite{Sampaio_NN2013,Tomasello_SREP2014,Xichao_NC2016,Xichao_PRB2016B,Xichao_PRB2022A,Xichao_PRB2022B,Zhang_NL2023}
\begin{equation}
\label{eq:energy-density}
\begin{split}
\varepsilon=&+A\left(\nabla\boldsymbol{m}\right)^{2}+D\left[\boldsymbol{m}\cdot\left(\nabla\times\boldsymbol{m}\right)\right] \\
&-\frac{M_{\text{S}}}{2}(\boldsymbol{m}\cdot\boldsymbol{B}_{\text{d}})-M_{\text{S}}(\boldsymbol{m}\cdot\boldsymbol{B}_{\text{a}}),
\end{split}
\end{equation}
where $A$ and $D$ stand for the ferromagnetic exchange and chiral exchange interaction constants, respectively.
$\boldsymbol{B}_{\text{d}}$ is the demagnetization field, and $\boldsymbol{B}_{\text{a}}$ denotes the applied magnetic field. 

The default parameters are based on the chiral bulk magnet Fe$_{0.5}$Co$_{0.5}$Ge, which are adopted from Ref.~\onlinecite{Yu_AM2024}:
$\gamma_{0}=2.211\times 10^{5}$ m A$^{-1}$ s$^{-1}$,
$\alpha=0.1-0.9$,
$M_{\text{S}}=239$ kA m$^{-1}$,
$A=4$ pJ m$^{-1}$,
and $D=-0.42$ mJ m$^{-2}$.
The helical period $L_{\text{0}}=4\pi\left|A/D\right|=120$ nm was confirmed experimentally by Lorentz transmission electron microscopy imaging~\cite{Yu_AM2024}.
The driving force provided by the spin-transfer torques is controlled by changing the applied current density $j$ with the simplified assumption of $P=1$, i.e., $u\sim j$.
All simulations are performed by using the \textsc{mumax$^3$} micromagnetic simulator~\cite{MuMax1,MuMax2} on NVIDIA GeForce RTX 3060 Ti and RTX 3070 graphics processing units.
The mesh size is set to $5$ $\times$ $5$ $\times$ $5$ nm$^3$ to ensure good computational accuracy and efficiency.

\vbox{}
%-%-%-%-%-%-%-%-%-%-%-%-%-%-%-%-%-%-%-%-%-%-%-%-%-%-%-%-%-%-%
\section{3 | RESULTS AND DISCUSSION}
\label{se:Results}
%-%-%-%-%-%-%-%-%-%-%-%-%-%-%-%-%-%-%-%-%-%-%-%-%-%-%-%-%-%-%

%-%-%-%-%-%-%-%-%-%-%-%-%-%-%-%-%-%-%-%-%-%-%-%-%-%-%-%-%-%-%
\subsection{3.1 | Initial \emph{metastable} bimeron states}
\label{se:Initial}
%-%-%-%-%-%-%-%-%-%-%-%-%-%-%-%-%-%-%-%-%-%-%-%-%-%-%-%-%-%-%

\noindent
We first study the relaxed static spin configurations in the chiral bulk magnet before applying the driving current.
We consider a randomly distributed spin configuration in the model with a default damping parameter of $\alpha=0.3$. The intrinsic magnetic parameters are given in Section~\blue{2}.
\emph{The relaxed state contains mainly in-plane spin structures, forming some irregular domain walls with certain out-of-plane magnetization (i.e., a helical state).}
It is a common state that can be found in chiral bulk magnets with zero magnetocrystalline anisotropy, because that the demagnetization effect (i.e., the shape anisotropy) favors the in-plane spin configurations.

We then apply an out-of-plane magnetic field to the model, where the magnetic field first increases from $B_z=0$ mT to $B_z=+400$ mT with a step change of $10$ mT, and then decreases from $B_z=+400$ mT to $B_z=0$ mT with the same step change. After each step change of the magnetic field, the system is fully relaxed and some \emph{metastable} topological spin textures are found at certain out-of-plane magnetic fields.
As shown in Figure~\ref{FIG1}(C), we focus on the \emph{metastable} spin textures obtained when decreasing the out-of-plane magnetic field from $B_z=+400$ mT to $B_z=0$ mT, where a transformation between skyrmions and bimerons are found.
It can be seen that when $B_z=+400$ mT, the background magnetization of the chiral bulk magnet is fully polarized along the $+z$ direction (i.e., the spin-up direction), while six chiral skyrmions with spin-down cores are found. These skyrmions are of Bloch-type due to the chiral exchange interactions in the bulk magnet. The total topological charge of the model $Q$ is thus equal to $-6$.

By decreasing the out-of-plane magnetic field to $B_z=0$ mT, the six skyrmions in the chiral bulk magnet smoothly transform into six bimerons, as a result of the conservation of topology.
In the absence of the out-of-plane magnetic anisotropy, the demagnetization effect in the bulk magnet favors an in-plane magnetized system, while the out-of-plane magnetic field leads to a perpendicularly magnetized system. Therefore, the competition between the out-of-plane magnetic field and demagnetization effect could result in the mutual transformation between skyrmions in a perpendicularly magnetized background and bimerons in an in-plane magnetized background, which has been confirmed in recent experiments~\cite{Yu_AM2024,Ohara_NL2022}.
It should be noted due to the demagnetization effect, the spin configurations are not uniform in the thickness dimension of the bulk magnet, as can be seen obviously in Figure~\ref{FIG1}(C) with $B_z=0-200$ mT.
The reason is that the demagnetization effect in the bulk magnet favors more in-plane magnetization components near the topmost and bottommost surfaces of the bulk magnet.

In the following, \emph{we focus on the relaxed metastable bimerons obtained at $B_z=+100$ mT during the decreasing of the out-of-plane magnetic field,} which we will use as the initial state in the simulation of the current-induced bimeron dynamics.
We will also discuss the effect of the out-of-plane magnetic field on the current-induced dynamics later in this paper.

\vbox{}
%-%-%-%-%-%-%-%-%-%-%-%-%-%-%-%-%-%-%-%-%-%-%-%-%-%-%-%-%-%-%
\subsection{3.2 | Bimeron proliferation and aggregation}
\label{se:Aggregation}
%-%-%-%-%-%-%-%-%-%-%-%-%-%-%-%-%-%-%-%-%-%-%-%-%-%-%-%-%-%-%

\noindent
We first focus on the dynamics of bimerons in the chiral bulk magnet driven by spin-transfer torques from an in-plane current in the presence of an out-of-plane magnetic field of $B_z=+100$ mT.
The relaxed initial spin configurations in the chiral bulk magnet are obtained by scanning the out-of-plane magnetic field as discussed above in Section~\blue{3.1}.
As shown in Figure~\ref{FIG1}(C), the initial state at $B_z=+100$ mT includes six relaxed bimerons with a total topological charge of $Q=-6$.
We then apply an in-plane current to drive the bimerons into motion, where we first assume a default damping parameter of $\alpha=0.3$ and a default driving current density of $j=0.9\times 10^{12}$ A m$^{-2}$. The current flows toward the right ($+x$) direction, i.e., the electrons flow toward the left ($-x$) direction. Other parameters are given in Section~\blue{2}.
The in-plane current is applied for $500$ ns from $t=0$ ns.
In Figure~\ref{FIG2}(A), it can be seen that when $\beta$ is larger than $0.5$ or equal to $0.1$, the absolute value of the total topological charge $|Q|$ in the current-driven system increases with time; however, $|Q|$ is independent of time when $\beta$ ranges between $0.2$ and $0.4$.

The increase of $|Q|$ indicates the creation of new topological spin textures induced by the driving current, while the time-independent $|Q|$ indicates the six bimerons given in the initial state are only driven into trivial translational motion (see \blue{Supplementary Video 1}).
Therefore, we focus on the non-trivial case with the parameters $\alpha=0.3$, $\beta=0.6$, $B_z=+100$ mT, and $j=0.9\times 10^{12}$ A m$^{-2}$, as indicated by the green dashed line in Figure~\ref{FIG2}(A), and explore the time-dependent spin configurations and energy.
As shown in Figure~\ref{FIG3}(A), the absolute total topological charge $|Q|$ in the chiral bulk magnet first increases from $6$ at $t=0$ ns to $38$ at $t=400$ ns, and then decreases to $32$ at $t=500$ ns; namely, the total topological charge of the final state is about five times larger than that of the initial state.
We find that the increase of $|Q|$ is due to the current-induced creation of new bimerons (see \blue{Supplementary Video 2}), as demonstrated in the time-dependent snapshots of the spin configurations in the chiral bulk magnet [see Figure~\ref{FIG4}(A)].

The rapid creation of new bimerons induced by the in-plane current leads to not only the proliferation but also the aggregation of bimerons in the chiral bulk magnet, as shown in Figure~\ref{FIG4}(B). Namely, most bimerons in the chiral bulk magnet form a cluster during their motion driven by the in-plane current. The aggregation effect is due to the attractive bimeron-bimeron interaction as well as the current-induced compression of bimerons.
The attractive interaction between bimerons favors the formation of bimeron chains~\cite{Zhang_PRB2020,Li_NJPCM2020}, while the compression leads to the shrink of bimerons and forms a compact arrangement of bimerons.
The formation of bimeron chains has also been observed experimentally in chiral bulk magnets~\cite{Yu_AM2024}.
However, the proliferation of bimerons during the current-induced motion of bimerons has not yet been uncovered explicitly in experiments.
We also note that the decrease of absolute total topological charge after $t=400$ ns in Figure~\ref{FIG3}(A) may be caused by an over aggregation of bimerons, because that the current-induced compression between topological spin textures could be destructive, usually leading to shrink and collapse of topological spin textures~\cite{Xichao_PRB2022A}.

In Figure~\ref{FIG3}(B), it can be seen that the total energy of the system increases generally to a higher level during the current-induced proliferation and aggregation of bimerons, which suggests that the dynamically aggregate bimerons are unstable and disequilibrium compared to the initial static state.
The time-dependent in-plane magnetization components of the system indicate that more spins in the chiral bulk magnet are reorientated from the $-x$ direction to the $-y$ direction [see Figure~\ref{FIG3}(C)].
The time-dependent out-of-plane magnetization component of the system indicates that some spins in the chiral bulk magnet are reorientated from the in-plane directions to the $+z$ direction [see Figure~\ref{FIG3}(D)], justifying the formation of new bimerons in the system as each bimeron has a net out-of-plane magnetization component due to its perpendicularly magnetized cores.

In Figure~\ref{FIG5}, we further show the micromagnetic energy terms of the system as functions of time during the current-induced proliferation and aggregation of bimerons.
Figure~\ref{FIG5}(A) shows that the ferromagnetic exchange energy increases with time, while the DM interaction energy decreases with time [see Figure~\ref{FIG5}(B)]. The reason is that the proliferation of bimerons is associated with the formation of more domain wall structures, which are favored by the DM interaction but unfavored by the ferromagnetic exchange interaction.
In principle, the ferromagnetic exchange interaction favors a uniformly magnetized ferromagnetic single-domain state, while the DM interaction is an antisymmetric exchange interaction that stabilizes domain wall structures.
The demagnetization energy of the system increases with time [see Figure~\ref{FIG5}(C)] as it favors more in-plane spin configurations but the current-induced new bimerons have cores with out-of-plane spins, either spin-up or spin-down [see Figure~\ref{FIG4}].
The Zeeman energy of the system decreases with time as the perpendicularly magnetized bimeron cores are favored by the applied out-of-plane magnetic field [see Figure~\ref{FIG5}(D)].
\emph{Here we also note that the time-dependent change of each energy term in Figure~\ref{FIG5} after $t=400$ ns shows generally opposite tendency compared to the stage before $t=400$ ns, which can be explained by the annihilation of bimerons, justifying the decrease in $Q$ after $t=400$ ns in Figure~\ref{FIG3}(a).}

In addition to the non-adiabatic spin-transfer torque strength $\beta$, we also find that the proliferation and aggregation of bimerons induced by the in-plane current depend on the value of the damping parameter $\alpha$. As indicated by the time-dependent total topological charge of the chiral bulk magnet in Figure~\ref{FIG2}(B), we find that the in-plane current will result in the proliferation and aggregation of bimerons only when $\alpha$ is smaller than $0.4$ for the system with $\beta=0.6$, $j=0.9\times 10^{12}$ A m$^{-2}$, and $B_z=+100$ mT.
On the other hand, for the focused case with $\alpha=0.3$, $\beta=0.6$, and $B_z=+100$ mT, we find that the current-induced proliferation and aggregation of bimerons could happen only when the applied driving current density is larger than $0.3\times 10^{12}$ A m$^{-2}$.
These results suggest that the strength of the non-adiabatic spin-transfer torque $\beta$ plays an important role in the proliferation of bimerons in the bulk magnet, while the motion of bimerons and the inter-bimeron interactions should, in principle, lead to the aggregation of bimerons.

As the Hall angle of the in-plane current-induced translational motion of topological spin textures is subject to the values of \emph{$|\beta-\alpha|$}, as reported previously in the skyrmion system~\cite{Zhang_IEEE2017}, the results given in \emph{Figure~\ref{FIG2}(A) and Figure~\ref{FIG2}(B)} also suggest that the system with a larger value of $|\beta-\alpha|$ could show the proliferation and aggregation of bimerons when the spin-transfer torques are strong enough to excite new bimerons, while driving the bimerons into motion with certain nonzero Hall angles.

\emph{In principle, both the bimeron and skyrmion carry integer topological charges and could have a Hall motion due to the topological Magnus force~\cite{Gobel_PRB2019}. However, the Hall motion of a bimeron is more complex than that of a skyrmion because the bimeron is not a localized spin texture and its orientation depends on the in-plane background magnetization. The Hall angle of a bimeron also depend on its shape and internal configuration in both the in-plane and out-of-plane dimensions. As a result, the bimerons with different orientations or configurations may show different Hall angles. On the other hand, as shown in Figure~\ref{FIG4}(A) and Supplementary Video 2, a few unstable bimerons with $Q=+1$ are created and stabilized dynamically during the proliferation stage, which could move along largely different paths compared to metastable bimerons with $Q=-1$. The collisional interactions between these bimerons are very effective as they are not moving along parallel paths, which is vital to the spawning of new bimerons.}

\emph{Namely, the nonzero and different bimeron Hall angles in the system ensure that the bimerons in the chiral bulk magnet are able to interact with each other effectively before forming dynamically stable aggregate states or bimeron chains. When an aggregate state is formed by metastable bimerons with $Q=-1$, the aggregate state moves with a certain Hall angle, and meanwhile, unstable bimerons with $Q=+1$ are diminished and annihilated eventually.}
\emph{For the same reason, the system with $|\beta-\alpha|=0$ is unable to show the proliferation and aggregation of bimerons as the intrinsic Hall angles for bimerons and skyrmions with $Q=\pm 1$ are equal to zero when $|\beta-\alpha|=0$. Consequently, bimerons with zero Hall angles will move along parallel paths and cannot interact with each other effectively to excite and create new bimerons.}

\vbox{}
%-%-%-%-%-%-%-%-%-%-%-%-%-%-%-%-%-%-%-%-%-%-%-%-%-%-%-%-%-%-%
\subsection{3.3 | Relaxation of aggregate bimerons}
\label{se:Relaxation}
%-%-%-%-%-%-%-%-%-%-%-%-%-%-%-%-%-%-%-%-%-%-%-%-%-%-%-%-%-%-%

\noindent
The current-induced rapid creation of a number of new bimerons is essential for the proliferation and aggregation of bimerons in the chiral bulk magnet. The creation of new bimerons \emph{may be due to} the fact that the bimeron lattice state is a more stable state in the chiral bulk magnet. This can be examined by relaxing the dynamically aggregate bimerons in the absence of the driving current.
Hence, we focus on the case with the default parameters $\alpha=0.3$, $\beta=0.6$, $B_z=+100$ mT, and $j=0.9\times 10^{12}$ A m$^{-2}$, as indicated by the green dashed lines in Figure~\ref{FIG2}, where an in-plane current is injected into the chiral bulk magnet for $500$ ns from $t=0$ ns.
We turn off the in-plane current at $t=500$ ns (i.e., $j=0$ A m$^{-2}$) and relax the system into an equilibrium stable or metastable state, as shown in Figure~\ref{FIG6} (see \blue{Supplementary Video 3}).

In Figure~\ref{FIG6}(A), we show the time-dependent absolute total topological charge of the system $|Q|$ for both the current-driven (i.e., $t=0-500$ ns) and current-free relaxation (i.e., $t=500-1000$ ns) periods.
It can be seen that $|Q|$ is independent of time during the relaxation, which indicates no annihilation of bimerons happens during the relaxation. However, the total energy of the system rapidly decreases to a stable value when the in-plane current is turned off [Figure~\ref{FIG6}(B)], suggesting the system is relaxed into a metastable state with a fixed number of bimerons in the chiral bulk magnet.
In particular, the total energy of the relaxed system after the current injection is smaller than that of the initial state at $t=0$ ns, which further indicates the relaxed state with $|Q|=32$ at $t=1000$ ns is actually more stable than that of the initial state with $|Q|=6$ at $t=0$ ns, although the initial state is also a metastable state.

In Figure~\ref{FIG6}(C), we show the time-dependent snapshots of the spin configurations during the current-free relaxation, where one can find that the aggregate bimerons are dispersed in the chiral bulk magnet and form a deformed bimeron lattice with a few defects in the lattice structure. Such a phenomenon suggests that the bimeron lattice structure are favored in the given system, and this is also the reason why the in-plane current is able to excite more bimerons.

It is worth mentioning that the spin configurations of the bimerons in the relaxed bimeron lattice is not uniform in the thickness dimension of the chiral bulk magnet due to the demagnetization effect, as can be seen from the snapshots at $t=1000$ ns in Figure~\ref{FIG6}(C).
The demagnetization effect in the chiral bulk magnet favors more in-plane magnetization near the topmost and bottommost surfaces, while the magnetization in the middle of the bulk magnet could have more out-of-plane component.
For this reason, it can be seen that the spin configurations in the middle of the bulk magnet are more like a special skyrmion lattice with compactly arranged but deformed skyrmions, while the spin configurations in the topmost and bottommost layers are more like a deformed bimeron lattice, forming a unique honeycomb structure with mixed topological spin textures in the thickness dimension.

\vbox{}
%-%-%-%-%-%-%-%-%-%-%-%-%-%-%-%-%-%-%-%-%-%-%-%-%-%-%-%-%-%-%
\subsection{3.4 | Magnetic field-dependent dynamics}
\label{se:Field}
%-%-%-%-%-%-%-%-%-%-%-%-%-%-%-%-%-%-%-%-%-%-%-%-%-%-%-%-%-%-%

\noindent
As the bimeron state can be found at different out-of-plane magnetic fields in the chiral bulk magnet, as pointed out in Section~\blue{3.1}. We further investigate the effect of the external out-of-plane magnetic field on the current-induced proliferation and aggregation dynamics of bimerons.
Here, we focus on the case with default parameters $\alpha=0.3$, $\beta=0.6$, and $j=0.9\times 10^{12}$ A m$^{-2}$. We apply an in-plane current to drive the system at different out-of-plane magnetic fields of $Bz=0-400$ mT. The in-plane current is injected into the chiral bulk magnet for $500$ ns from $t=0$ ns, where the initial state at $t=0$ ns is a relaxed state obtained at different out-of-plane magnetic fields (see Section~\blue{3.1}).

Interestingly, we find that the proliferation and aggregation of bimerons can be induced by the in-plane current only when the applied external out-of-plane magnetic field is of an appropriate magnitude, mainly within the range $B_z=50-150$ mT, which is indicated by the time-dependent $|Q|$ in Figure~\ref{FIG7}.
In Figure~\ref{FIG8}, we show the corresponding time-dependent spin configurations in the chiral bulk magnet during the application of the in-plane current. The initial states at $B_z=0-400$ mT include only six relaxed bimerons or skyrmions at $t=0$ ns [see Figure~\ref{FIG8}(A)]. 
It can be seen obviously that the proliferation and aggregation of bimerons happen in the systems with an out-of-plane magnetic field of $B_z=50-150$ mT. The current-induced proliferation and aggregation of bimerons are especially prominent when $B_z=100-150$ mT, where one can find many dynamically aggregate bimerons at the final state at $t=500$ ns [see Figure~\ref{FIG8}(B)].

It should be noted that skyrmions could also be stabilized by out-of-plane magnetic fields in the given system; however, they do not demonstrate the current-induced proliferation. Also, in the absence of the out-of-plane magnetic field (i.e., $B_z=0$ mT), the system only favors some in-plane spin configurations; consequently, some bimerons with out-of-plane cores given in the initial state are annihilated during their motion driven by the in-plane current, and no new bimerons are created.
Therefore, it can be seen that the external magnetic field applied in the out-of-plane direction could enhance the stability of the bimerons in the chiral bulk magnet as bimerons have core spins aligned in the out-of-plane directions.
On the other hand, the external out-of-plane magnetic field with an appropriate magnitude could foster the current-induced proliferation of bimerons, and thus, enabling the current-induced aggregation of bimerons.
However, a strong out-of-plane magnetic field may result in the transformation of bimerons to skyrmions, and skyrmions will be driven into translational motion by the in-plane current.
Namely, no new skyrmions will be created during the current-induced motion of skyrmions in the given system.

\vbox{}
%-%-%-%-%-%-%-%-%-%-%-%-%-%-%-%-%-%-%-%-%-%-%-%-%-%-%-%-%-%-%
\section{4 | CONCLUSIONS}
\label{se:Conclusions}
%-%-%-%-%-%-%-%-%-%-%-%-%-%-%-%-%-%-%-%-%-%-%-%-%-%-%-%-%-%-%

\noindent
In conclusion, we computationally found the proliferation and aggregation dynamics of bimerons in a chiral bulk magnet induced by the in-plane current in the presence of an out-of-plane magnetic field.
\emph{The difference between the damping parameter and the non-adiabatic spin-transfer torque strength as well as the applied current density play important roles in the proliferation and subsequent aggregation of bimerons.}
The external magnetic field applied in the out-of-plane direction can also control the bimeron proliferation and aggregation.
Namely, in principle, one can turn on and off the proliferation and aggregation dynamics by adjusting the magnitude of the out-of-plane magnetic field in a dynamic way. It is also possible to control the proliferation and aggregation dynamics by tuning the applied current density, provided that the spin-transfer torques and damping parameters are appropriate.

We also demonstrated that the aggregate bimerons induced by the in-plane current can relax into a special deformed honeycomb bimeron lattice with a few lattice structure defects when the in-plane current is turned off.
This provides an effective way to create a large number of bimerons and construct bimeron lattice structures in chiral magnets, which has the potential to lead to promising non-conventional computation based on controllable bimeron lattices or aggregate bimerons.

\emph{We note that the quenched disorder in the bimeron-hosting system is an important factor to consider in the future. In principle, the quenched disorder could lead to more complex motion of topological spin textures and may also affect the lattice formation. By introducing the quenched disorder with randomly distributed pinning sites, for example, one could reduce the bimeron Hall angle due to the creep motion, and therefore, suppress the creation of new bimerons.}

Our results also suggest a new platform for studying any aggregation-induced effects in condensed matter systems, including but not limited to the aggregation-induced lattice formation, aggregation-induced lattice deformation, aggregation-induced phase transitions in bimeron solids, and aggregation-induced spin wave emission.
Our results may motivate the development of new aggregate science that enhances our understanding of dynamic phenomena in chiral magnets.

%-%-%-%-%-%-%-%-%-%-%-%-%-%-%-%-%-%-%-%-%-%-%-%-%-%-%-%-%-%-%
\vbox{}
\noindent\textbf{ACKNOWLEDGEMENTS}
%-%-%-%-%-%-%-%-%-%-%-%-%-%-%-%-%-%-%-%-%-%-%-%-%-%-%-%-%-%-%

\noindent
X.Z., X.Y., and M.M. acknowledge support by CREST, the Japan Science and Technology Agency (Grant No. JPMJCR20T1).
\emph{M.M. also acknowledges support by the Grants-in-Aid for Scientific Research from JSPS KAKENHI (Grants No. JP20H00337, No. JP23H04522, and JP24H02231), and the Waseda University Grant for Special Research Projects (Grant No. 2024C-153).}
Y.Z. acknowledges support by the Shenzhen Fundamental Research Fund (Grant No. JCYJ20210324120213037), the Guangdong Basic and Applied Basic Research Foundation (Grant No. 2021B1515120047), the Shenzhen Peacock Group Plan (Grant No. KQTD20180413181702403), the National Natural Science Foundation of China (Grant No. 12374123), and the 2023 SZSTI Stable Support Scheme.
%-%-%-%-%-%-%-%-%-%-%-%-%-%-%-%-%-%-%-%-%-%-%-%-%-%-%-%-%-%-%

%-%-%-%-%-%-%-%-%-%-%-%-%-%-%-%-%-%-%-%-%-%-%-%-%-%-%-%-%-%-%
\vbox{}
\noindent\textbf{CONFLICT OF INTEREST STATEMENT}
%-%-%-%-%-%-%-%-%-%-%-%-%-%-%-%-%-%-%-%-%-%-%-%-%-%-%-%-%-%-%

\noindent
The authors declare no conflicts of interest.
%-%-%-%-%-%-%-%-%-%-%-%-%-%-%-%-%-%-%-%-%-%-%-%-%-%-%-%-%-%-%

%-%-%-%-%-%-%-%-%-%-%-%-%-%-%-%-%-%-%-%-%-%-%-%-%-%-%-%-%-%-%
\vbox{}
\noindent\textbf{DATA AVAILABILITY STATEMENT}
%-%-%-%-%-%-%-%-%-%-%-%-%-%-%-%-%-%-%-%-%-%-%-%-%-%-%-%-%-%-%

\noindent
The data that support the findings of this study are available from the corresponding authors upon reasonable request.
The micromagnetic simulator \textsc{mumax$^3$} used in this work is publicly accessible at http://mumax.github.io/index.html.
%-%-%-%-%-%-%-%-%-%-%-%-%-%-%-%-%-%-%-%-%-%-%-%-%-%-%-%-%-%-%

\vbox{}
%-%-%-%-%-%-%-%-%-%-%-%-%-%-%-%-%-%-%-%-%-%-%-%-%-%-%-%-%-%-%

%-%-%-%-%-%-%-%-%-%-%-%-%-%-%-%-%-%-%-%-%-%-%-%-%-%-%-%-%-%-%

%-%-%-%-%-%-%-%-%-%-%-%-%-%-%-%-%-%-%-%-%-%-%-%-%-%-%-%-%-%-%
\vbox{}
\noindent\textbf{SUPPORTING INFORMATION}
%-%-%-%-%-%-%-%-%-%-%-%-%-%-%-%-%-%-%-%-%-%-%-%-%-%-%-%-%-%-%

\noindent
Additional supporting information can be found online in the Supporting Information section at the end of this article.
%-%-%-%-%-%-%-%-%-%-%-%-%-%-%-%-%-%-%-%-%-%-%-%-%-%-%-%-%-%-%

%%%%%%%%%%%%%%%%%%%%%%%%%%%%%%%%%%%%%%%%%%%%%%%%%%%%%%%%%%%%
\end{document}